\documentclass[aps,prc,twocolumn,showpacs,superscriptaddress,floatfix,10pt]{revtex4-1}
\usepackage{graphicx}
\usepackage{dcolumn}
\usepackage{bm}
\usepackage{amsfonts}
\usepackage{amssymb}
\usepackage{amsmath}

\begin{document}

\title{Confronting measured near and sub-barrier fusion cross-sections for $^{20}$O+$^{12}$C with a microscopic method}

\author{R. T. deSouza}
\author{S. Hudan}
\affiliation{Department of Chemistry and Center for Exploration of Energy and 
Matter, Indiana University, Bloomington, Indiana 47405, USA}
\author{V.E. Oberacker}
\author{A.S. Umar}
\affiliation{Department of Physics and Astronomy, Vanderbilt University, Nashville, Tennessee 37235, USA}

\date{\today}


\begin{abstract}
Recently measured fusion cross-sections
for the neutron-rich system $^{20}$O+$^{12}$C are compared to
dynamic, microscopic calculations using time-dependent density functional theory. The calculations
are carried out on a
three-dimensional lattice and performed both with and without a constraint on the density.
The method has no adjustable parameters, and its only input is the
Skyrme effective NN interaction. While the microscopic DC-TDHF calculations lie closer to
the experimental data than standard fusion systematics they underpredict the experimental data significantly.
\end{abstract}
\maketitle

The outer crust of an accreting neutron star provides a unique environment in which nuclear
reactions can occur.
It has been proposed that the fusion of two neutron-rich light nuclei in the outer crust could provide a heat source to ignite
thermonuclear fusion of $^{12}$C~+~$^{12}$C and produce a signature X-ray superburst~\cite{Horowitz08}.
To date, however, a limited amount is known either experimentally or theoretically about the fusion of
neutron-rich nuclei. Pioneering experiments with heavy nuclei indicate that the fusion
below the barrier may be enhanced \cite{Loveland06,Liang12}. Such an enhancement has recently been associated with the
importance of neutron transfer channels which effectively lowers the fusion barrier~\cite{UO12}.
In the case of fusion of two neutron-rich {\it light} nuclei (Z$<$20), even less is known. In principle, this is the most
promising domain as neutron-rich nuclei up to the drip line can be experimentally produced.

Recent experimental measurement of near-barrier fusion in the system $^{20}$O~+~$^{12}$C~\cite{Rudolph12}
suggests that the
fusion cross-section is enhanced relative to the predictions of the
Bass model~\cite{Bass77}.
As the empirical Bass model
is based upon the systematics of known fusion cross-sections near $\beta$-stability,
it does not include the increased
importance of neutron transfer channels for neutron-rich nuclei.
The aim of this
paper is to directly compare the experimental results with a microscopic approach,
namely the
time dependent Hartree-Fock (TDHF)
theory.

The experiment was performed at the SPIRAL1 facility at the GANIL accelerator complex in Caen,
France. An $^{20}$O beam with an
intensity of $1-2 \times 10^4$ p/s impinged on a $100$~$\mu$g/cm$^2$ thick $^{12}$C target. The energy of
the beam on target was varied between 1~MeV/A and 2~MeV/A in order to measure the fusion
excitation function.
Experimental details have been previously published \cite{Rudolph12} and are summarized here only for completeness.
Nuclei produced by fusion subsequently de-excite via evaporation of neutrons and light charged
particles (Z$\le$2) forming evaporation residues. These residues were detected in two
segmented silicon detectors located downstream of
the target and identified by measuring both their energy and time-of-flight.
The annular detectors spanned the angular range
$3.54^{\circ}\le\theta_{lab} \le21.8^\circ$.
Due to the presence of a large atomic background in the experiment,
a coincidence between an emitted charged particle and the evaporation residue
was necessary to distinguish fusion reactions.
Statistical model calculations with a Hauser-Feshbach model, evapOR,
indicate that, depending on the excitation energy of the compound nucleus formed,
approximately 15-25~$\%$ of fusion
reactions de-excite via emission of at least one charged particle.

The time-dependent Hartree-Fock (TDHF) theory provides a useful foundation for a
fully microscopic many-body theory of large amplitude collective
motion. It is therefore well suited to describing deep-inelastic and fusion reactions~\cite{Negele82,Simenel11}.
Only in recent years has it become feasible to perform TDHF calculations on a
3D Cartesian grid without any symmetry restrictions
and with much more accurate numerical methods~\cite{Simenel11,US91,UO06,Simenel12,GM08,DD-TDHF}.
In addition, the quality of effective interactions has been substantially
improved~\cite{CB98,Klu09a,KL10,USR}. TDHF theory predicts an energy
density functional which is determined by the given effective NN interaction.
One may therefore view TDHF as a special case of a time-dependent
density functional theory (TDDFT), a concept used in many areas of
nuclear physics, condensed-matter physics, and chemistry.

Over the past several years,
the Density Constrained
Time-Dependent Hartree-Fock (DC-TDHF) method for calculating
heavy-ion potentials~\cite{UO06a} was utilized to calculate fusion cross-sections. We have applied this method
to calculate fusion and capture cross-sections above and below the barrier to about $20$ systems to date,
examples of which can be found in Refs.~\cite{UO06d,UO07a,UO09b,UO10a,KU12}.
Recently, we have also investigated sub-barrier fusion between
nuclei that occur in the neutron star crust~\cite{UO12}. In all cases, we have found
good agreement between the measured fusion cross-sections and the DC-TDHF results.
This agreement is rather remarkable given the fact that the only input in DC-TDHF is the
Skyrme effective N-N interaction, and there are no adjustable parameters.

The TDHF equations for the single-particle wave functions
\begin{equation}
h(\{\phi_{\mu}\}) \ \phi_{\lambda} (r,t) = i \hbar \frac{\partial}{\partial t} \phi_{\lambda} (r,t)
            \ \ \ \ (\lambda = 1,...,A) \ ,
\label{eq:TDHF}
\end{equation}
can be derived from a variational principle~\cite{Negele82}.
In the present TDHF calculations we use the Skyrme SLy4 interaction~\cite{CB98} for the nucleons
including all of the time-odd terms in the mean-field Hamiltonian~\cite{UO06}.
The numerical calculations are carried out on a 3D Cartesian lattice. For
the calculations shown in this work,
the lattice spans $40$~fm along the collision axis and $24-30$~fm in
the other two directions, depending on the impact parameter. We first generate very accurate
static HF wave functions for the two nuclei on the 3D grid. In the second
step, we apply a boost operator to the single-particle wave functions. The time-propagation
is carried out using a Taylor series expansion (up to orders $10-12$) of the unitary mean-field propagator,
with a time step $\Delta t = 0.4$~fm/c.

\begin{figure}[!htb]
\includegraphics*[width=8.6cm]{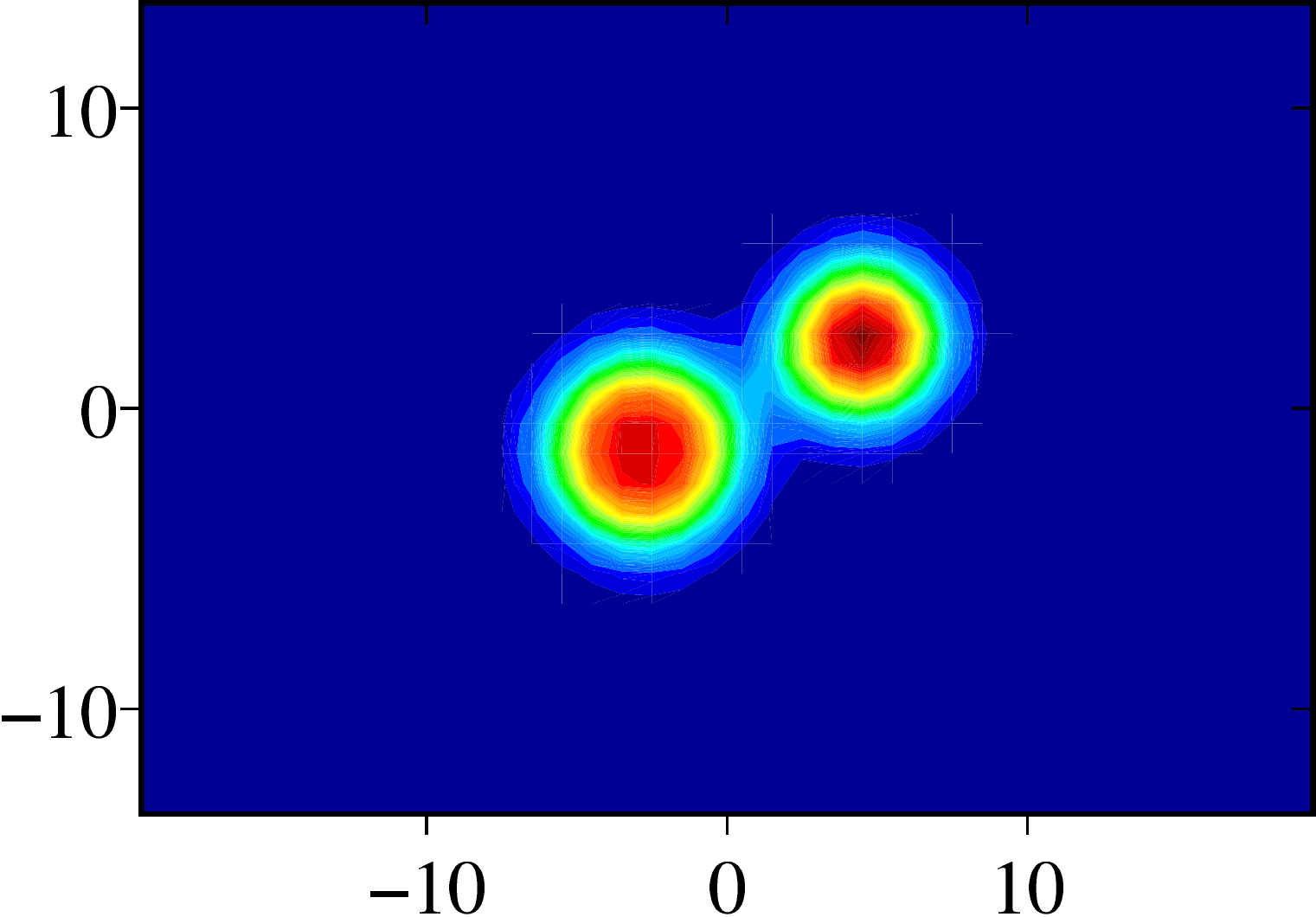}
\caption{(Color online) Unrestricted TDHF+BCS calculation for $^{20}$O+$^{12}$C
at $E_{\mathrm{c.m.}}=9.5$~MeV and impact parameter $b=2.5$~fm.
Shown are mass density contours shortly after a neck has formed between the
two fragments.}
\label{fig1}
\end{figure}

Presented in Fig.~\ref{fig1} is a contour plot of the
mass density during a collision which clearly shows the formation of a neck
between the two fragments.
This density distribution, shown here for $^{20}$O+$^{12}$C at $E_{\mathrm{c.m.}}=9.5$~MeV,
is representative of collisions for similar systems.
As the collision proceeds in the TDHF calculation, transport of protons and neutrons
between the two nuclei can be followed within the theory. For larger impact parameters the larger angular momentum of the system leads the two nuclei
to separate and a deep-inelastic reaction occurs. For smaller impact parameters the disrupting
influence of angular momentum and Coulomb repulsion is insufficient to overcome the nuclear
attraction and fusion results. By examining the density distribution as the two nuclei
fuse into one within the calculation, one clearly observes the occurrence of a damped
dipole resonance and surface waves. Deep inelastic and fusion reactions
are the dominant reaction channels in this energy domain. Distinguishing between these two
types of reactions is realized by examining the density distribution as a function of time and
observing whether one or two large fragments result from the collision.

In the absence of a true quantum many-body theory of barrier tunneling,
all current sub-barrier fusion calculations assume the existence of an ion-ion potential
$V(R)$ which depends on the internuclear distance $R$. Most of the theoretical
fusion studies are carried out with the coupled-channels (CC) method~\cite{HW07,Das07,IH09,EJ10}
in which one uses empirical ion-ion potentials (typically Woods-Saxon potentials,
or double-folding potentials with frozen nuclear densities). In contrast to
the use of these empirical potentials we have adopted a microscopic approach
to extract heavy-ion interaction potentials
$V(R)$ from the TDHF time-evolution of the dinuclear system which
describes the dynamics of the underlying nuclear shell structure.

In the DC-TDHF approach~\cite{UO06a},
the TDHF time-evolution proceeds uninhibitedly. At certain times $t$ or,
equivalently, at certain internuclear distances $R(t)$ the instantaneous TDHF density
is used to perform a static Hartree-Fock energy minimization while constraining the
proton and neutron densities to be equal to the instantaneous TDHF densities.
This means that we allow the single-particle wave functions to rearrange themselves in such a way
that the total energy is minimized, subject to the TDHF density constraint.
In a typical DC-TDHF run, we utilize a few thousand time steps, and the density
constraint is applied every $10-20$ time steps.
We refer to the minimized energy as the ``density constrained energy'' $E_{\mathrm{DC}}(R)$.
The ion-ion interaction potential $V(R)$ is obtained by subtracting the constant binding energies
$E_{\mathrm{A_{1}}}$ and $E_{\mathrm{A_{2}}}$ of the two individual nuclei
\begin{equation}
V(R)=E_{\mathrm{DC}}(R)-E_{\mathrm{A_{1}}}-E_{\mathrm{A_{2}}}\ .
\label{eq:vr}
\end{equation}

In direct TDHF calculations the fusion cross-section is calculated by determining the maximum impact parameter for
which fusion occurs and applying the sharp cut-off approximation. For example,
in the case of the reaction $^{20}$O+$^{12}$C
at $E_{\mathrm{c.m.}}=9.5$~MeV we find that impact parameters $b \leq b_{max}=4.075$~fm result
in fusion, while impact parameters $b > b_{max}$ lead to deep-inelastic reactions.
Using the sharp cut-off model, the fusion cross-section is given by
$\sigma_{\mathrm{fus}} = \pi b_{max}^2 = 52.2$~fm$^2 = 522$~mb.
In contrast, in DC-TDHF method fusion cross-sections are obtained by integrating the
Schr\"odinger equation for the potential $V(R)$ with a coordinate-dependent mass~\cite{UO09b}.

\begin{figure}[!htb]
\includegraphics*[width=8.6cm]{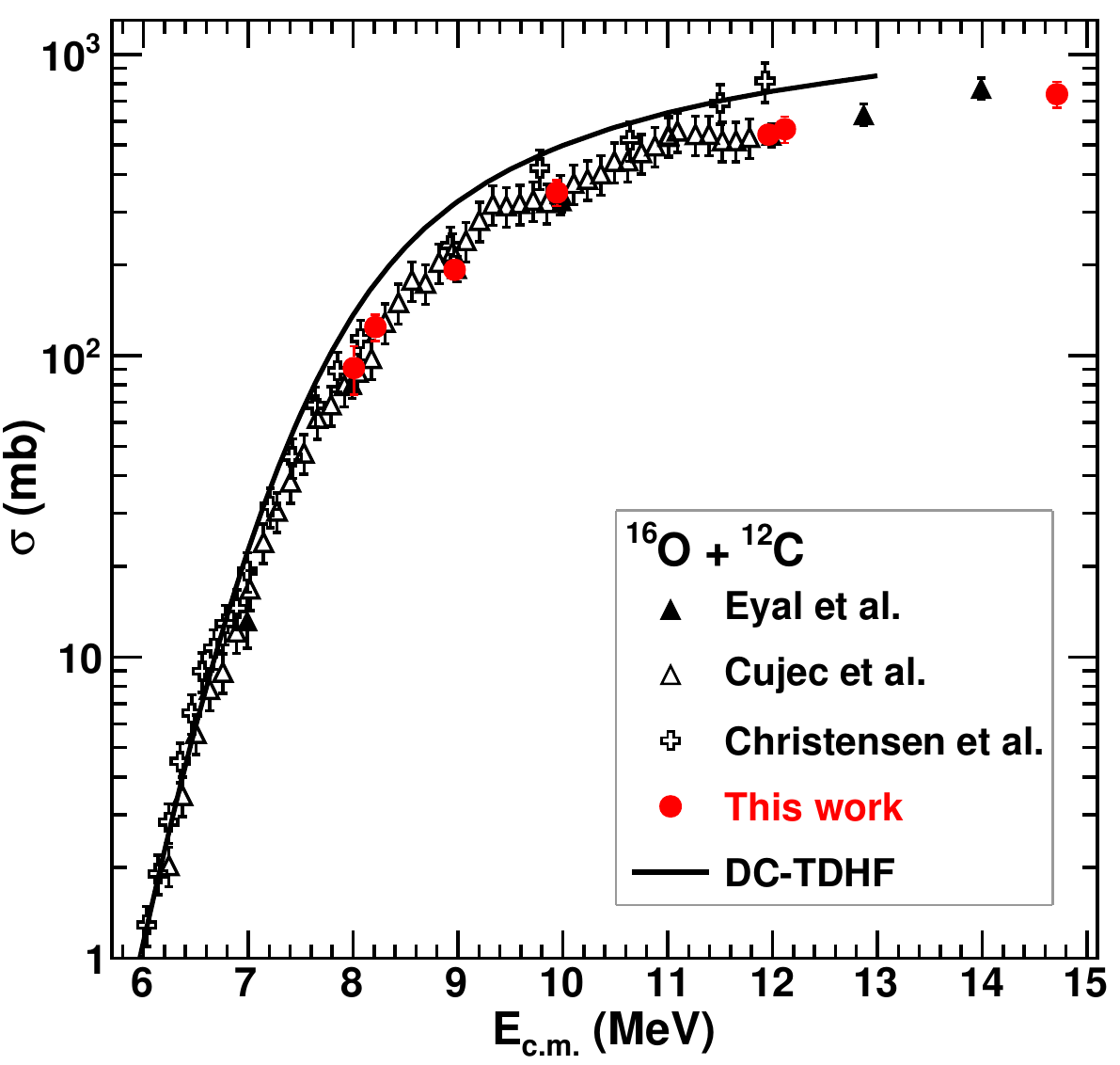}
\caption{(Color online) Comparison of the experimentally measured fusion excitation
function
with the predictions of the DC-TDHF method for $^{16}$O+$^{12}$C.
Data from \cite{Cujec76,Eyal76,Christensen77} shown.}
\label{fig2}
\end{figure}

We begin the comparison of the microscopic calculations with experimental data by examining
the well studied system $^{16}$O+$^{12}$C. Shown in Fig.~\ref{fig2} are four sets of
experimental data for the total fusion cross-section along with the corresponding microscopic DC-TDHF calculations.
The experimental
techniques used to determine the fusion cross-section range from gamma spectroscopy of
the charged particle channels (Christensen and Cujec) to direct measurement of the
evaporation residues (Eyal and this work).
It should be stressed that the data represented by the red filled circles (this work) utilized the
same experimental technique and setup as for the $^{20}$O+$^{12}$C results subsequently
presented.
For energies
$E_{\mathrm{c.m.}}<9.5$~MeV all the experimental data are in agreement. At higher energies
however,
the cross-sections measured by Christensen \textit{et al.} slightly exceed that of the three other datasets.
We have no explanation at present for the larger cross-sections measured by Christensen \textit{et al.}
Since the cross-section measurements of Eyal, Cujec, and the present work are all in good
agreement we take these cross-sections to accurately represent the true fusion
cross-section. It
is interesting to note that the DC-TDHF calculations also slightly exceed the measured
cross-sections for $E_{\mathrm{c.m.}}>7$~MeV. At the largest energies measured this excess
is of the order of 20~$\%$.
Indeed, the agreement of DC-TDHF results with the data of Christensen \textit{et al.} at
the highest energies was somewhat surprising since TDHF dynamics for light heavy-ions
at these energies
do not properly account for various breakup channels present for these systems, and
results in a fusion-like composite system with long-time collective oscillations.
In coupled-channel calculations this discrepancy is cured by introducing a small imaginary
potential in the vicinity of the potential minimum~\cite{EJ10}.
Having established the degree of confidence through the comparison of the fusion
cross-sections in
$^{16}$O+$^{12}$C, we then calculated
fusion in $^{20}$O+$^{12}$C.

While for isolated $^{16}$O and $^{12}$C nuclei the Hartree-Fock (HF) ground state is
found to be spherical in agreement with both theory and experiment, for $^{20}$O the
HF calculations predict a prolate
quadrupole deformation of $\beta_2=0.25$. This
deformation is in disagreement with self-consistent mean
field calculations with pairing (Skyrme-HFB)~\cite{DS04} which predict a spherical
nucleus. Moreover,
the measured energy level spectrum~\cite{Tilley98} also shows this nucleus to be spherical.
In addition, a measurement of the magnetic moment of $^{20}$O~\cite{Gerber76}
also indicates its spherical nature.
We attribute this prolate deformation predicted by the HF calculations
to the lack of pairing in the method.
One unfortunate consequence of the DC-TDHF approach in calculating the fusion cross-section
lies in the treatment of pairing during the collision process. In unrestricted
TDHF calculations, the BCS occupation numbers can be kept frozen during the collision to
have correct initial states.
This approximation cannot be utilized in the DC-TDHF method because
the static HF solution coupled with a constraint on the instantaneous TDHF density
for the combined system requires the reevaluation of the occupation numbers for the
lowest energy solution. Consequently, calculations with the DC-TDHF method
do not include pairing. On qualitative grounds, it can be argued that this omission
should result in a slightly larger
prediction of the fusion cross-section. The reason is that pairing results in a spherical
$^{20}$O nucleus, and the fusion barrier for a spherical nucleus is higher than the lowest
barrier for a deformed nucleus.
In order to calculate the fusion cross-section for this system within DC-TDHF method
we therefore take an average of all initial orientation angles
$\beta$ of the deformed $^{20}$O nucleus, where $\beta$ is defined as the angle
between the internuclear distance vector and the symmetry axis of the deformed
nucleus.

\begin{figure}[!htb]
\includegraphics*[width=8.6cm]{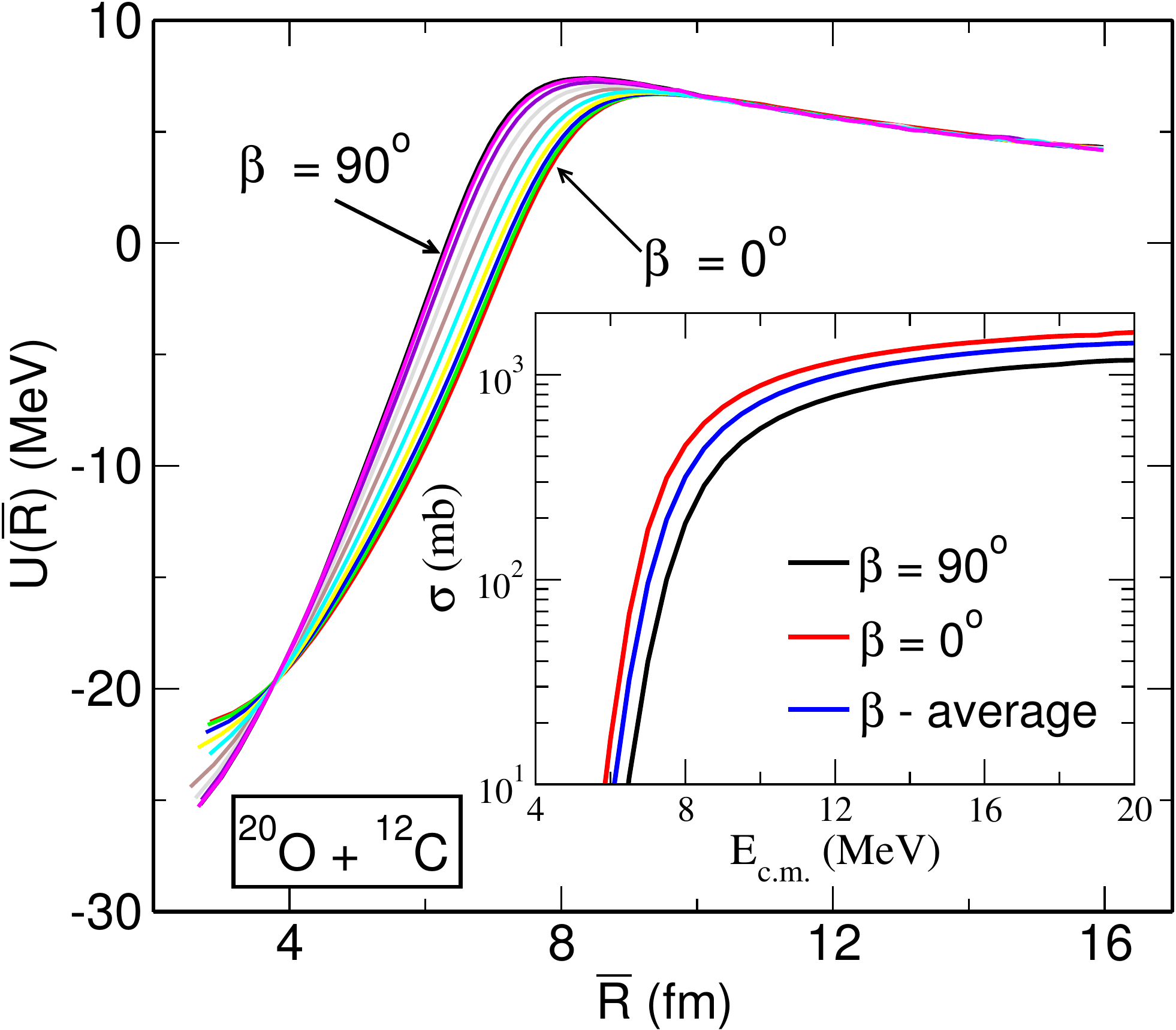}
\caption{(Color online) Transformed heavy-ion potentials $U(\bar{R})$, corresponding to the
constant reduced mass $\mu$. The potentials have been calculated for a series
of initial orientation angles $\beta$ of the deformed $^{20}$O nucleus (no pairing). Shown in the
inset is the total predicted cross-section as a function of $E_{\mathrm{c.m.}}$
for $\beta$=0$^\circ$, $\beta$=90$^\circ$, as well as the cross-section
averaged over $\beta$.}
\label{fig3}
\end{figure}

As the collision occurs, using TDHF dynamics, it is possible to compute
the corresponding coordinate
dependent mass parameter $M(R)$~\cite{UO09b}.
At large distance $R$, the mass $M(R)$ is equal to the
reduced mass $\mu$ of the system. At smaller distances, when the nuclei overlap, the
mass parameter generally increases.
In order to calculate the fusion cross-section more easily, one
can replace the coordinate-dependent mass $M(R)$ and the original potential $V(R)$
with the constant mass $\mu$ and the ``transformed potential'' $U(\bar{R})$,
using a scale transformation~\cite{UO09b}.
In Fig.~\ref{fig3} we display the transformed potentials $U(\bar{R})$ for
initial orientation angles $\beta = 0^{\circ},10^{\circ},...,90^{\circ}$
of $^{20}$O.
For sufficiently large separation between the two nuclei, R $\ge$9, fm all the
transformed heavy-ion potentials are the same regardless of the orientation of
the two nuclei. With decreasing distance as the two nuclei come into contact, the
heavy-ion potentials differ.
For a nucleus with prolate deformation, the orientation angle $\beta = 0^{\circ}$
leads to the lowest potential barrier. This reduction in the potential occurs
because the distance between the
nuclear surfaces is minimized for this orientation which is important because of
the short-range
nature of the strong interaction. The Coulomb interaction is accurately
calculated by solving the 3D Poisson equation numerically during the collision.
We observe that with increasing orientation
angle, the barrier height increases and the barrier position is shifted
to smaller distances.

Displayed in the inset of Fig.~\ref{fig3} is the fusion
excitation function associated with the orientations of $\beta$=0$^\circ$
and 90$^\circ$ as well as the angle averaged result. As one might qualitatively
expect, based upon the heavy-ion potentials,
for a given $E_{\mathrm{c.m.}}$,
the orientation $\beta = 0^{\circ}$ with the lowest potential barrier
corresponds to the largest cross-section.

\begin{figure}[!htb]
\includegraphics[width=8.6cm]{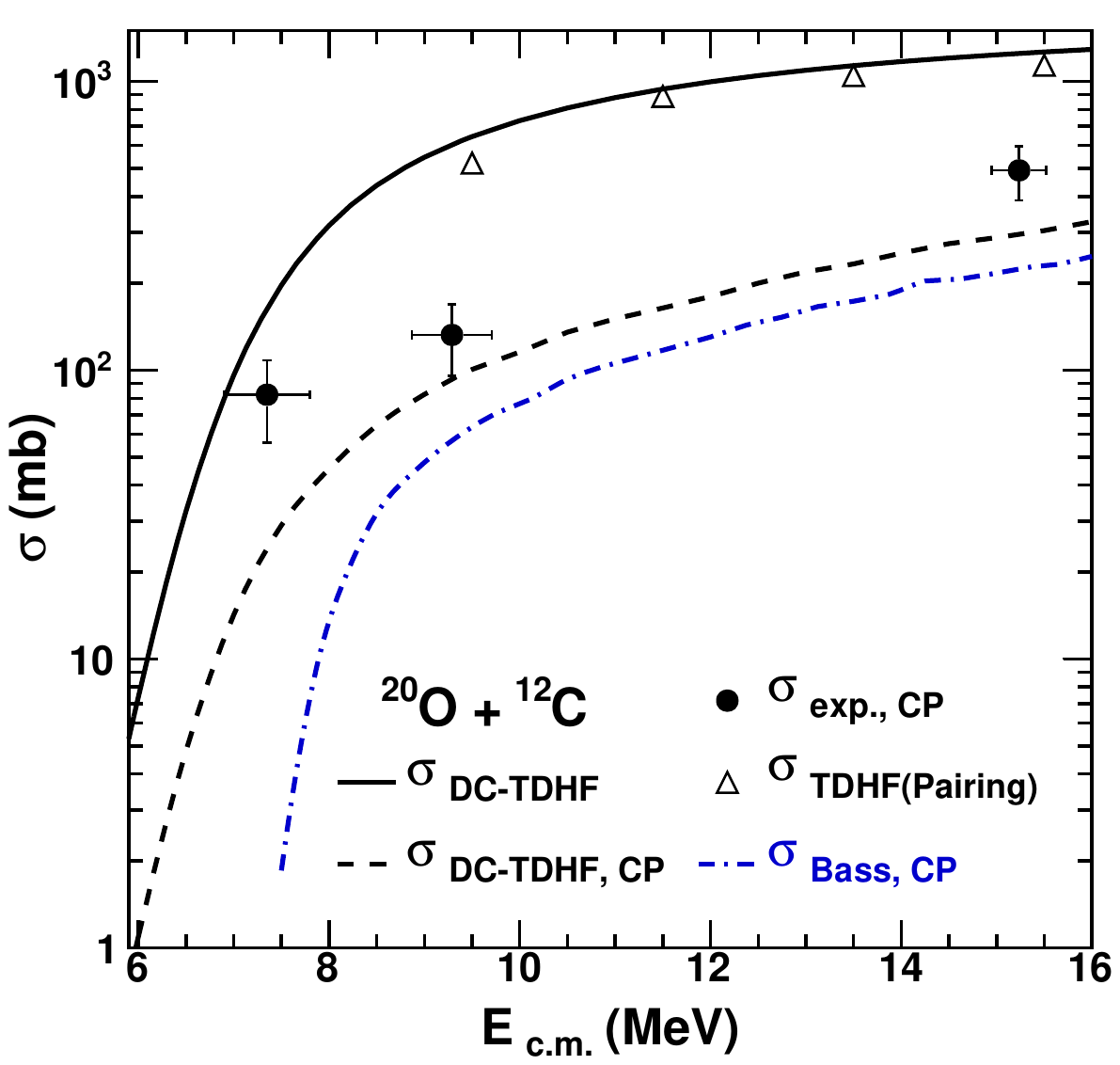}
\caption{(Color online) Comparison of the angle-averaged DC-TDHF fusion cross-sections
with experimental data for $^{20}$O+$^{12}$C.}
\label{fig4}
\end{figure}

In Fig.~\ref{fig4} we compare the microscopic calculations using
the DC-TDHF method to the experimental data. The experimental data at
$E_{\mathrm{c.m.}}$= 7.35, 9.29, and 15.24~MeV are shown as the filled circles.
Due to the atomic background previously
mentioned only the fraction of the fusion cross-section associated with subsequent
charged particle emission was experimentally measured.
In order to compare the microscopic calculations with the experimental data, we
therefore calculated the fraction of compound nuclei produced at each incident energy that
de-excite via emission of at least one charged particle. To calculate the
de-excitation of the fused nuclei we utilized a Hauser-Feshbach statistical model, evapOR.
The theoretically predicted fusion cross-section associated with subsequent charged
particle channels is depicted by the dashed line. This predicted cross-section clearly
underpredicts the
experimentally measured cross-section. At the highest energy,
$E_{\mathrm{c.m.}}$=15.24~MeV, the predicted cross-section is 60~$\%$ of the
experimentally measured one while at the lowest energy the predicted cross-section is
substantially lower, only 30~$\%$ of the experimental value. For reference, we also
present, as a dot-dash line, the fusion cross-section predicted by the Bass systematics that is associated
with charged particle emission. This cross-section is less than that of the
DC-TDHF method most likely reflecting the influence of neutron transfer channels in
aiding the fusion process.
It should be appreciated that not only does the DC-TDHF method predict a larger
cross-section at all energies as compared to the Bass systematics, but this increase
grows with decreasing $E_{\mathrm{c.m.}}$. This result suggests that the neutron transfer
becomes more important in the sub-barrier domain.
The result that the DC-TDHF method underpredicts the
experimental data by a significant amount is noteworthy. Moreover, it should be noted that
the lack of
pairing in the DC-TDHF method and the resulting deformation of the $^{20}$O,
as previously discussed, acts to increase the predicted cross-section implying that the
discrepancy between theoretical prediction and experimental data is at least as large as
that evident in Fig.~\ref{fig4}.

In order to assess the impact of pairing on the measured
cross-sections more quantitatively we have performed unrestricted TDHF calculations
(no barrier tunneling) at energies
above the barrier. These calculations
were initialized with BCS/Lipkin-Nogami pairing for $^{20}$O which resulted in a
spherical nucleus, consistent with the experimental data.
During the collision of the $^{20}$O with the $^{12}$C
the BCS occupation numbers are kept frozen. The results of these calculations are
presented as the open triangles in Fig.~\ref{fig4}. A slight reduction in the total
cross-section is evident. This reduction is of the order of 5-20~$\%$ with the largest
reduction for the lowest energy point calculated. From these calculations one can
infer that the inclusion of pairing in the DC-TDHF calculations should result in a
slight reduction of the predicted cross-section, thus increasing the discrepancy with
the experimental data as anticipated.

It is tantalizing to speculate about possible reasons for the discrepancy between
the theoretical predictions and the experimental data.
In general terms, the discrepancy between the
experimental data for the charged particle channels in Fig.~\ref{fig4}
(solid points) and the corresponding DC-TDHF calculations can be thought of as
originating either from an underprediction of the total fusion cross-section or from
an underestimation of the relative importance of the charged particle decay in the
de-excitation of the compound nucleus. Either or both of these sources could explain
the underprediction of the cross-section. It is therefore important to not only
measure the total
fusion cross-section but also the cross-section for individual decay channels.
Furthermore, since the underprediction
exists for energies well above the barrier, $E\approx 9.5$~MeV,
it is not simply a question of enhanced
tunneling. Moreover, beyond the overall underprediction of the
microscopic method, one observes that the experimental data manifests a
slower fall-off for the measured cross-section with decreasing
incident energy as compared to the DC-TDHF calculations. This experimentally determined
energy dependence is highly provocative and it remains to be seen whether the total fusion
cross-section also exhibits this slower fall-off. As this fall-off is intimately related to the
neutron transfer channels, near and sub-barrier fusion of neutron-rich nuclei
provides direct access to
the extent of the neutron density distribution.


\begin{acknowledgments}
This work has been supported by the U.S. Department of Energy under Grant No.
DE-FG02-96ER40975 (Vanderbilt University) and DE-FG02-88ER-40404 (IU).
\end{acknowledgments}


\bibliography{paper}

\end{document}